\documentclass[letterpaper,twocolumn,10pt]{article}
\usepackage{usenix2019_v3}

\usepackage{tikz}
\usepackage{amsmath}

\usepackage{filecontents}

\usepackage{amssymb}
\usepackage{array}
\usepackage{booktabs}
\usepackage{fullpage}
\usepackage{threeparttable}
\usepackage{wasysym}
\usepackage{flushend}
\usepackage{caption}
\usepackage{subcaption}
\usepackage{longtable}
\usepackage{scrtime}
\usepackage[multiple]{footmisc}
\usepackage{footnote}
\usepackage{makecell}
\usepackage{multirow}
\usepackage{rotating}
\usepackage{subcaption}
\usepackage{multicol}
\usepackage{graphicx}
\usepackage{float}
\usepackage{listings}
\usepackage{cleveref}
\usepackage{textcomp}
\usepackage{footnote}
\usepackage{enumitem}
\usepackage{lipsum}
\usepackage{mathtools}

\crefformat{section}{\S#2#1#3} % see manual of cleveref, section 8.2.1
\crefformat{subsection}{\S#2#1#3}
\crefformat{subsubsection}{\S#2#1#3}

\usepackage{epsfig,hyperref,color,graphicx,xspace,pifont,dirtree}

\usetikzlibrary{shapes.geometric, arrows}
\usepackage{pgfplotstable}

\usepackage{pgfplots}
    \pgfplotsset{compat=1.3}
    
    \definecolor{bblue}{HTML}{4F81BD}
    \definecolor{rred}{HTML}{C0504D}
    \definecolor{ggreen}{HTML}{9BBB59}

\newcommand{\rom}[1]{{\em\lowercase\expandafter{(\romannumeral #1\relax)}}}
\newcommand{\nom}[1]{{\em\lowercase\expandafter{(#1\relax)}}}

\newcommand\cve[1]{{\href{https://cve.mitre.org/cgi-bin/cvename.cgi?name=CVE-#1}{CVE-#1}}}

\lstdefinestyle{customc}{
  belowcaptionskip=1\baselineskip,
  breaklines=true,
  frame=ltrb,
  xleftmargin=\parindent,
  language=C,
  showstringspaces=false,
  basicstyle=\footnotesize\ttfamily\bfseries,
  keywordstyle=\bfseries\color{green!60!black},
  commentstyle=\itshape\color{purple!60!black},
  identifierstyle=\color{blue},
  stringstyle=\color{red},
}
\lstdefinestyle{customasm}{
  belowcaptionskip=1\baselineskip,
  frame=R,
  xleftmargin=\parindent,
  language=[x86masm]Assembler,
  basicstyle=\footnotesize\ttfamily\bfseries,
  commentstyle=\itshape\color{purple!60!black},
}

\newcommand\chk{\ding{51}}

\begin{document}
%-------------------------------------------------------------------------------

%don't want date printed
\date{}

% make title bold and 14 pt font (Latex default is non-bold, 16 pt)
%\title{\Large \bf \textmu Tiles: Flexible In-Process Least Privilege Enforcement of Memory Units }
\title{\Large \bf \textmu Tiles: Efficient Intra-Process Privilege Enforcement of Memory Regions}

%for single author (just remove % characters)
\author{
{\rm Zahra Tarkhani}\\
University of Cambridge
\and
{\rm Anil Madhavapeddy}\\
University of Cambridge
% copy the following lines to add more authors
% \and
% {\rm Name}\\
%Name Institution
} % end author

%\author{{\rm Anonymous Authors}}
\maketitle

%-------------------------------------------------------------------------------
\begin{abstract}
%------------------------------------------------------------------------------
With the alarming rate of security advisories and privacy concerns on connected devices, there is an urgent need for strong isolation guarantees in resource-constrained devices that demand very lightweight solutions. However, the status quo is that Unix-like operating systems do not offer privilege separation inside a process. 
Lack of practical fine-grained compartmentalization inside a shared address space leads to private data leakage through applications' untrusted dependencies and compromised threads.  To this end, we propose \textmu Tiles, a lightweight kernel abstraction and set of security primitives based on mutual distrust for intra-process privilege separation, memory protection, and secure multithreading. \textmu Tiles takes advantage of hardware support for virtual memory tagging (e.g., ARM memory domains) to achieve significant performance gain while eliminating various hardware limitations. Our results (based on OpenSSL, the Apache HTTP server, and LevelDB) show that \textmu Tiles is extremely lightweight (adds $\approx 10KB$ to kernel image) for IoT use cases. It adds negligible runtime overhead ($\approx 0.5\%-3.5\%$) and is easy to integrate with existing applications for providing strong privilege separation.  
  
\end{abstract}

\section{Introduction}\label{intro}

\begin{center}

\begin{figure}[t]
\includegraphics[width=\linewidth,height=4.5cm]{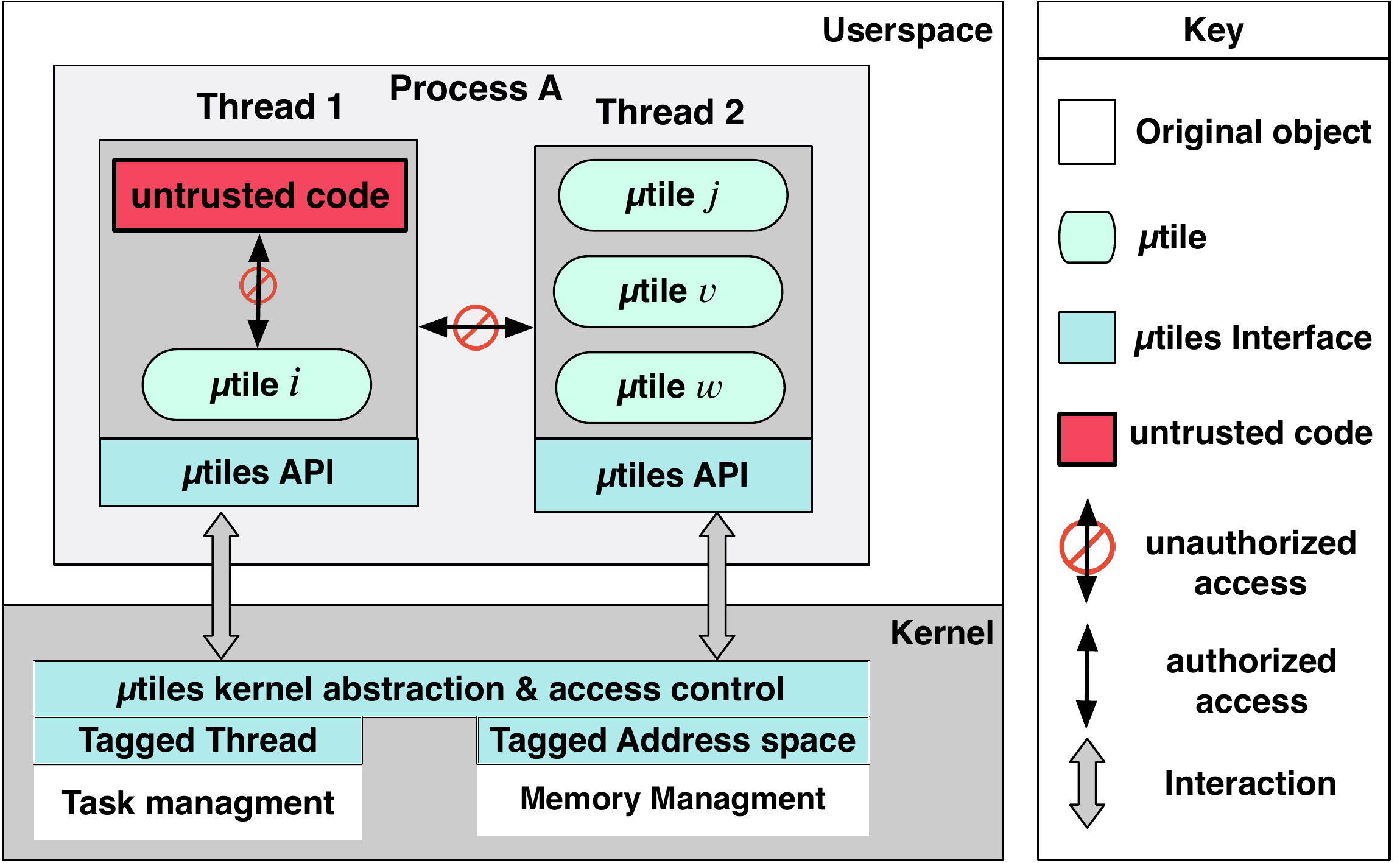}

\caption{High-level architecture of \textmu Tiles: it provides strong intra-process isolation and privilege separation. Each thread can define its own trust boundaries in the form of \textmu Tiles that are protected memory regions. \textmu Tiles are guarded against both untrusted code within the same thread as well any untrusted threads.}
\label{hlevel}
\end{figure}

 \end{center}

\begin{table}[htb]
\resizebox{\columnwidth}{!}{%
  \footnotesize
  \begin{tabular}{|c|l|l|c|}
      \hline

    & \bf CVE       & \bf Description  & \bf uTiles \\
    \hline
    \parbox[t]{2mm}{\multirow{7}{*}{\rotatebox[origin=c]{90}{\bf In-Process threats}}}
    
    & \cve{2019-9345} & Shared mapping bug  &\chk \\
    & \cve{2019-9423} & missing bounds check      & \chk \\
    & \cve{2019-15295} & unsafe third party library      & \chk \\
    & \cve{2019-1278} &  unsafe third party library   &  \chk\\
    & \cve{2018-0487} & unsafe third party library    & \chk \\
    
    & \cve{2017-1000376} &  unsafe native bindings     & \chk \\

    & \cve{2014-0160} &  Heartbleed bug     & \chk \\ 
  
    \hline
    
        \parbox[t]{2mm}{\multirow{3}{*}{\rotatebox[origin=c]{90}{\bf Other}}}

                &  && \\

     & \cve{2018-0497} & SW side-channels  &   \\
      & \cve{2017-5754} & HW side-channels & \\

    \hline

  \end{tabular}}
  \caption{\label{t:cve-table}A representative selection of vulnerabilities that cause sensitive content leakage. The attacks with a tick can be mitigated by using {\bf \textmu Tiles} protection.}
  \label{cvetab}
\end{table}

Many software attacks target sensitive content in an application's address space, usually through remote exploits, malicious third-party libraries, or unsafe language vulnerabilities.  
Conventional operating systems consider processes as units of isolation. However, particularly in IoT use cases, most applications generate and analyze highly sensitive data in a single process for efficiency reasons. 
This leads to real threats (summarised in Table~\ref{t:cve-table}) that require effective protection against:\\
\rom{1} an application's secret data (e.g., private keys or user passwords) can be leaked in the presence of compromised third-party libraries like OpenSSL~\cite{durumeric2014matter};
\rom{2} privileged functions can be misused to access private content~\cite{deng2015iris};
\rom{3} applications written in memory-safe languages such as Rust or OCaml are vulnerable via unsafe external libraries that jeopardizes all other safety guarantees~\cite{almohri2018fidelius,lamowski2017sandcrust};
and
\rom{4} in multithreaded servers attackers can exploit vulnerabilities (e.g., buffer overflows) so the compromised thread can access sensitive data owned by other threads~\cite{tserver}.
This whole class of attacks could be avoided by providing a practical way to enforce the least privilege within a shared address space.  

Process-based isolation is the primary compartmentalization technique
for security-sensitive applications such as OpenSSH to separate their components into different processes~\cite{provos2003preventing,nsjail,firejail}. However, this usually causes a large overhead and requires redesigning an application from scratch using a multiprocess architecture (e.g., Chrome) that is impractical for most multithreaded applications such as web servers. 
Previous work such as Privtrans~\cite{brumley2004privtrans} and Wedge~\cite{bittau2008wedge} provide automatic process-based isolation of applications with a huge overhead ($\approx 80\%-40$x slowdown). Such systems' impracticality is partially inherited from the conventional process abstractions such as \texttt{fork} that already suffers from various efficiency and security issues~\cite{baumann2019fork}. Even \texttt{fork} alternatives such as \texttt{clone}, are not flexible enough for fine-grained data sharing between processes for security-critical resources. 
We need a better abstraction for shrinking the trust boundaries from inter-process to intra-process; so developers can effectively prevent in-process attacks and build secure multithreaded applications. 

The importance of these security threats results in significant improvement in hardware support for efficient memory isolation~\cite{mte,mpk,arm2012architecture,watson2015cheri}. 
However, simple APIs for utilizing such hardware features are not effective due to the complexity of attacks as well as various hardware limitations~\cite{vahldiek2019erim,park2018libmpk}. We need a more principled mitigation approach. In particular, the requirements of real-world IoT applications show a practicality gap in the existing solutions (summarized in Table~\ref{compare}) that need to be covered.

\begin{table}[t]

\footnotesize
\newcommand*\rot[1]{\hbox to1em{\hss\rotatebox[origin=br]{-30}{#1}}}
\newcommand*\feature[1]{\ifcase#1 -\or\LEFTcircle\or\CIRCLE\fi}
\newcommand*\f[8]{\feature#1&\feature#2&\feature#3&\feature#4&\feature#5&\feature#6&\feature#7&\feature#8&\fcontinued}
\newcommand*\fcontinued[1]{\feature#1}
\makeatletter
\newcommand*\ex[4]{#1\tnote{#2}&#3&%
    \f#4&\@firstofone
}
\makeatother
\newcolumntype{G}{ccc}
\begin{threeparttable}
\resizebox{\columnwidth}{!}{%

\begin{tabular}{@{}lc GG !{\kern.1em} GG !{\kern.1em} GG@{}}
\toprule
Isolation mechanism &   & \multicolumn{8}{c}{Features} \\
\midrule
% rotated items
 && \rot{No compiler modification}
  & \rot{Low hardware dependency}
  & \rot{Flexible security policy}
 & \rot{Simplicity/Usability}
  & \rot{Low overhead}
& \rot{Unlimited isolated units}
& \rot{Multithreaded privilege separation}
 & \rot{POSIX compatibility}
 & \rot{Embedded device suitable}
 \\
\midrule
\ex{SFI/HFI~\cite{sehr2010adapting,zhou2014armlock,chen2016shreds}}{}{}  {010102021}\\
\ex{Tagged-VMA/TLB~\cite{litton2016light,vahldiek2019erim,park2018libmpk}}{}{}  {200112120}\\
\ex{LibOS\cite{porter2011rethinking,madhavapeddy2013unikernels}}{}{}  {120110011}\\
\ex{Process-based isolation\cite{brumley2004privtrans,bittau2008wedge}}{}{}  {220100021}\\
\ex{Capability hardware\cite{watson2015cheri,zeldovich2008hardware}}{}{}  {102112220}\\
\ex{DIFC-OSes~\cite{zeldovich2006making,krohn2007information}}{}{}  {222102110}\\

\ex{\textmu Tiles}{}{}  {212222222}\\

\bottomrule
\end{tabular}}
\begin{tablenotes}
\item $\feature2$ { }{ }{ } has the feature
\item $\feature1$ { }{ }{ } partially has the feature
\item $\text{{ }\feature0}$ { }{ }{ } does not have the feature
\end{tablenotes}   
\end{threeparttable}
\caption{{Overview of in-process isolation techniques. We consider these metrics for our design, focusing on the requirements of IoT use cases. }}
\label{compare}

\end{table}

In this paper, we present \textmu Tiles, a new OS abstraction for enforcing least privilege between threads and on slabs of memory within the same address space. Unlike previous work, \textmu Tiles security model allows each thread to selectively protect or share its memory compartments both from the untrusted code within itself as well as from any untrusted thread (see Figure~\ref{hlevel}). \textmu Tiles' access control layer maps threads' security policies to \textmu Tiles dedicated virtual memory (VM) abstraction. 
This VM manager provides an efficient memory tagging layer by bypassing most of the kernel's paging abstraction. It utilizes hardware-enforced VMA tagging (e.g., ARM memory domains~\cite{arm2012architecture} or Intel MPK~\cite{mpk}) to achieve low overhead. It should be noted that these hardware features have various security and practicality limitations (\S\ref{back}) that are mitigated by \textmu Tiles high-level abstraction. Our contributions can be summarised as follows:

\begin{itemize}
\item present a new kernel's security primitives based on mutual-distrust for intra-process privilege separation. It provides strong protection of private content, a secure multithreading model, and guarded communication within a shared address space. 
\item describe how to utilize modern CPU facilities for efficient memory tagging to avoid the overhead of existing solutions (due to TLB flushes, per-thread page tables, or nested page table management) while relieving the hardware limitations.
\item show that the solution is ultra-lightweight ($\approx 5K$ LoC) to be practical for embedded devices with a minimal memory footprint.
\item evaluate our implementation using real-world software such as  Apache HTTP server, OpenSSL, and Google's LevelDB, which shows  
\textmu Tiles add negligible runtime overhead for lightly modified applications while significantly improve their security by strong compartmentalization.
\end{itemize}

The remainder of this paper elaborates on the hardware features we use (\S\ref{back}), describes the architecture (\S\ref{over}) and implementation of \textmu Tiles (\S\ref{imp}), presents an evaluation (\S\ref{eval}) and finally the trade-offs of our approach (\S\ref{diss}).

\section{Goals \& Assumptions}\label{back}
The \textmu Tiles abstraction aims to enforce thread-granularity least privilege for memory accesses via the following principles and assumptions on the underlying hardware.

\subsection{Design Principles}\label{dp}

\textbf{Fine-grained strong isolation:}  All threads of execution should be able to define their security policies and trust models to selectively protect their sensitive resources. Current OS security models of sharing (``everything-or-nothing'') are not flexible enough for defining fine-grained trust boundaries within processes or threads (lightweight processes).\\
\textbf{Performance:} \textmu Tiles operations, including lunching, running, changing access permissions, and sharing across threads, should have minimal overhead. Moreover, untrusted (i.e., \textmu Tiles-independent) parts of applications should not suffer any overhead.\\
\textbf{Efficiency:} \textmu Tiles should be lightweight to be practical for embedded devices running on a few megabytes of memory and slow ARM CPUs.\\
\textbf{Compatibility:} It is difficult to provide strong security guarantees with no code modifications, and \textmu Tiles is no exception. We move most of these modifications into the Linux kernel (increasingly popular for embedded deployments~\cite{iot19}) and provide simple userspace interfaces. \textmu Tiles should be implemented without extensive changes to the Linux and not depend on a specific programming language, so existing applications can be ported easily.

To achieve effective isolation, we need a security model based on mutual-distrust that lets each thread protect its own \textmu Tiles from untrusted parts of the same thread as well as other threads and processes.
 Simply providing POSIX memory management within \textmu Tiles (e.g. \texttt{malloc} or \texttt{mprotect}) is inadequate. As a simple example, attackers can misuse the API for changing the memory layout of other threads' \textmu Tiles or unauthorized memory accesses. 
The \textmu Tiles interface needs~\rom{1} to provide isolation within a single thread;~\rom{2} to be flexible for sharing between threads, and ~\rom{3} to restrict all unauthorized permission changes or memory mappings modification of allocated \textmu Tiles. Previous work such as ERIM~\cite{vahldiek2019erim} or libMPK~\cite{park2018libmpk}
does not offer such security guarantees since their focus is more on performance and virtualization of the hardware protection keys.\\
We derive inspiration from Decentralised Information Flow Control (DIFC)~\cite{krohn2007information} but with a more constrained interface -- by not supporting information flow within a program, we avoid the complexities and performance overheads that typically involves.
 Existing DIFC kernels such as HiStar~\cite{zeldovich2006making} achieve our isolation goals, but requires a non-POSIX-based OS that makes it impractical for many applications, particularly IoT use cases. To have a practical and lightweight solution, we therefore built \textmu Tiles by modifying the Linux kernel and additionally utilizing modern hardware facilities for VMA tagging to deliver low overhead.

\begin{table*}[htb]
  \centering
  \footnotesize
  \resizebox{\textwidth}{!}{%
    \begin{tabular}{|l|l|l|}
    \hline
    Feature                   & ARM Memory Domains                             & Intel MPK                                                            \\ \hline
    Per process domains       & 16                                             & 16                                                                   \\ \hline
    Access control register   & DACR  (2 bits per domain, privileged register) & PKRU (2 bits per domain, userspace register)                         \\ \hline
    Access rights             & No-access, Full access, MMU default            & No-access, write-disable, MMU default                                \\ \hline
    Paging modes              & 2-level paging (bits 8:5, level 1 entries)     &  4-level paging (bits 62:59 of PDPTE)                                \\ \hline
    Address space privilege   &  Privileged \& usersapce                       &  Userspace only                                                      \\ \hline
    Specific page fault       & Domain fault                                   & PK fault                                                             \\ \hline
    Kernel virtual memory API & No support                                     & Limited support (\texttt{pkey\_mprotect}, \texttt{pkey\_alloc}, \texttt{pkey\_free}, and \texttt{mmap}) \\ \hline
    \end{tabular}}

    \caption{ARM memory domains vs Intel MPK: Despite being efficient building blocks of isolation, such features have various limitations that \textmu Tiles abstraction resolves to provide effective intra-process privilege separation.}
    \label{mpk}
\end{table*}

\begin{table}[htb]
\centering
\resizebox{\columnwidth}{!}{%

    \begin{tabular}{|l|l|l|}
    
    \hline
    Mode & Bits & Description                                                                     \\ \hline
    No Access   & 00 &  Any access causes a domain fault.                          \\\hline
    Manager     & 11 & Full accesses with no permissions check.\\\hline
    Client      & 01 &  Accesses are checked against the page tables   \\\hline
    Reserved    & 10 &   Unknown behaviour. \\ \hline
    \end{tabular}}
    \caption{ARM memory domains access permissions }
    \label{domains}
\end{table}

\subsection{HW-enforced VMA Tagging}\label{memdomain}

Modern CPUs have supported memory protection mechanisms that are more efficient than traditional paging. For example, VMA tagging features such as Intel Memory Protection Keys~\cite{mpk} and ARMv7 Memory Domains (MDs)~\cite{arm2012architecture} provide fast isolation by reducing page table walks and TLB flushes. Though the implementations across Intel and ARM vary considerably (see Table~\ref{mpk}), \textmu Tiles high-level VM abstraction can securely utilize these efficient building blocks while hiding their limitations.  In this paper, we primarily describe the design of \textmu Tiles for ARMv7-A that is a widely used CPU in IoT and mobile devices. Also, ARM-MD is a less flexible and more challenging interface to support (\S\ref{challenges}) that covers most MPK limitations as well.

As a summary of ARMv7-A memory management, page table entries consist of a virtual base address, a physical base address, Address Space Identifier (ASID) tags, domain IDs, and a set of flags for access control and other page attributes. It supports a two-level hierarchical page table when using a short-descriptor translation table format, and supports variable page sizes (1GB, 1MB, 64KB, and 4KB).
ARM supports two page tables simultaneously, using the hardware registers TTBR0 and TTBR1. A virtual address is mapped to a physical address by the CPU, depending on settings in TTBRC. This control register has a field that sets a split point in the address space. Addresses below the cutoff value are mapped through the page tables pointed to by TTBR0 (used per process), and addresses above the cutoff value are mapped through TTBR1 (used by the kernel). \\
The translation tables hold a four-bits \textbf{domain ID} ranging from $D0$ to $D15$. Access control for each domain is handled by setting a domain access control register (DACR) in $CP15$, which is a 32-bit register only accessible in the privileged processor modes. Each domain is assigned two bits in DACR, which defines its access rights. \\
The four possible access rights for a domain are No Access, Manager, Client, and Reserved (see Table~\ref{domains}). Those fields let the processor ~\rom{1} prohibit access to the domain mapped memory--No Access;~\rom{2} allow unlimited access to the memory despite permission bits in the page table-- Manager; or~\rom{3} let the access right be the same as the page table permissions--Client. Any access violation causes a domain fault. Changes to the DACR are low cost and activated without affecting the TLB. Hence changing domain permissions does not require TLB flushes.

\subsection{Challenges of Utilizing ARM-MD}\label{challenges}
Though ARM memory domains are a promising primitive in concept, the current hardware implementation and OS support suffer from significant problems that have prevented their broader adoption:\\
\noindent\textbf{Scalability:} ARM relies on a 32-bit DACR register and so supports only up to 16 domains. Allocating a larger register (e.g., 512 bits) would mean larger page table entries or additional storage for domain IDs.\\
\textbf{Flexibility:}
Unlike Intel MPK, ARM-MDs only apply to first-level entries; the second-level entries inherit the same permissions. This prevents arbitrary granularity of memory protections to small page boundaries and reduces the performance of some applications~\cite{cox2017efficient}.
Also, the DACR access control options do not directly mark a domain as read-only, write-only, or exec-only. So the higher-level VM abstraction should resolve these issues.\\
\textbf{Performance:} Changing the DACR is a fast but privileged operation, so any change of domain access permissions from userspace require a system call. This is unlike Intel MPK that makes its Protection Key Rights Register (PKRU) accessible directly from userspace.\\
\textbf{Userspace:} There is no Linux userspace interface for using ARM-MD; it is only used within the kernel to map the kernel and userspace into separate domains. In contrast, Linux already provides some basic support for utilizing Intel MPK from userspace.\\
\textbf{Security:} Though the DACR is only accessible in privileged mode, any syscall that changes this register is a potential breach that could cause the attacker to gain full control\footnote{An occasion that has happened once already through the misuse of the put\_user/get\_user kernel API (CVE-2013-6282)}. Also, since only 16 domains are supported, it is trivial to guess other domains' identifiers, making it essential to not expose these directly to application code.

\section{\textmu Tiles}\label{over}

We now describe \textmu Tiles architecture, which is a kernel abstraction for intra-process privilege separation with an emphasis on strong isolation, performance, and practicality for IoT use cases. \textmu Tiles abstraction contains three primary kernel's components for ~\nom{1} access control and least privilege enforcement,~\nom{2} threading and task management,~\nom{3} and dedicated virtual memory manager.

\subsection{Threat Model}

This paper focuses on two types of threats. First, memory-corruption based threats inside a shared address space that lead to sensitive information leakage; these threats can be caused by bugs or malicious third-party libraries (see Table~\ref{cvetab}). Second, attacks from threads that could get compromised by exploiting logical bugs or vulnerabilities (e.g., buffer overflow attacks, code injection, or ROP attacks). We assume the attacker can control a thread in a vulnerable multithreaded application, allocate memory, and spawn more threads up to resource limits by the OS and hardware. The attacker will try to escalate privileges through the attacker-controlled threads or gain control of another thread, e.g., by manipulating another thread's data or via code injection. The adversary may also bypass protection regions by exploiting race conditions between threads or by leveraging confused-deputy attacks.\\
\textmu Tiles thus provides isolation in two stages:~\nom{1} within a single thread (through utile\_lock/unlock calls), and~\nom{2} across threads in the same process. We consider threads to be security principals that define their security policies based on mutual-distrust within the shared address space. We protect each thread's \textmu Tiles against unauthorized, accidental, and malicious access or disclosure. Therefore, the TCB consists of the OS kernel, which performs security policy enforcement. It also assumes developers correctly specify their policies through the userspace interface for managing \textmu Tiles.\\
\textmu Tiles are not protected against covert channels based on shared hardware resources (e.g., a cache). Systems such as Nickel~\cite{sigurbjarnarson2018nickel} or hardware-assisted platforms such as Hyperflow~\cite{ferraiuolo2018hyperflow} could be a helpful future addition for side-channel protection on \textmu Tiles.

\subsection{Access Control Mechanism}\label{labeling}
Our modified Linux kernel enforces the principle of least privilege via a dynamic security policy based on DIFC~\cite{zeldovich2006making, wang2015between} and a simpler version of the Flume~\cite{krohn2007information} labeling with ~\textit{only two kernel objects} that are thread and address space.
Any thread $t$ has one secrecy(${SL_{t}}$) and integrity label (${IL_{t}}$) that each is set of unique tags. \textmu Tile objects (e.i., contiguous units of memory) have only one secrecy label instead of both types. The integrity violations are restricted in the higher-level by controlling the flow of threads labels; this improves performance and reduces complexity.\\
Privileges are represented in forms of two capabilities $\theta^{+}$  and $\theta^{-}$ per tag $\theta$ for adding or removing tags to/from labels. 
These capabilities are stored in a capability list $C_{p}$ per thread $p$.
Unique tags are assigned internally by the kernel by calling \texttt{utile\_create}. For improving security, none of \textmu Tiles API propagates tags in the userspace; all APIs access control is done internally within the kernel. The kernel allows secrecy information flow from $\alpha$ to $\beta$ only if $SL_{\alpha}\subseteq SL_{\beta}$, and integrity flow if $IL_{\beta}\subseteq IL_{\alpha}$. 
Every thread $p$ may change its label from $L_{i}$ to $L_{j}$ if it has the capability to add tags present in $L_{j}$ but not in $L_{i}$, and can drop the tags that are in $L_{i}$ but not in $L_{j}$. This is formally declared as ($L_{j}-L_{i}\subseteq C_{p}^{+}) \land (L_{i}-L_{j}\subseteq C_{p}^{-} )$.

When a thread has $\theta^{+}$ capability for \textmu Tile $\theta$, it gains the privilege to only access \textmu Tile $\theta$ \textit{with only the permission set by its owner} (read/write/execute).
The access privileges to each \textmu Tile can be different; hence, two threads can share a \textmu Tile, but the access privileges can differ. 
 Having a $\theta^{-}$ capability lets a thread to declassify \textmu Tile $\theta$. The declassification allows the thread to modify the \textmu Tile memory layout (by adding/removing pages to it), changing permissions, or copying the content to untrusted sources. Unsafe operations like declassifying \textmu Tiles or by endorsing a \textmu Tile as high-integrity require the thread to be an owner or an authority (\texttt{acts-for} relationship); which can be managed by \texttt{utile\_grant} and \texttt{utile\_revoke} calls (see Table~\ref{syscalls}).

\begin{table}[htb]
\resizebox{\columnwidth}{!}{%

    \begin{tabular}{|l|l|}
    \hline
    syscalls                        & Description                                                                                                \\ \hline
    utile\_transfer\_caps($u\_info*$,$tid$)         & passing only plus capabilities to thread $tid$\\ \hline
    utile\_declassify($u\_info*$)           & thread declassification or endorsement   \\ \hline
    utile\_grant($u\_info*$, $tid$)          & adds an acts-for or a delegation link to another thread \\ \hline
    utile\_revoke\_grant($u\_info*$, $tid$)    & removes an acts-for or a delegation link \\ \hline
    utile\_lock ($u\_info*$)         & disables access to set of  \textmu Tiles  \\ \hline
    utile\_unlock ($u\_info*$) & enables access to locked \textmu Tiles \\ \hline
    utile\_clone ($u\_info*$,int(*fn)(void*)...)$\to tid$      & creates a thread     \\ \hline

    \end{tabular}}
    
\caption{\textmu Tiles access control system calls. $tid$ represents a thread ID, \texttt{struct $u\_info*$} is the owner list of \textmu Tiles IDs and other fields for ownership management and capabilities per \textmu Tile. There is no direct propagation of labels that are security-critical data structures, and security policies are enforced within the kernel.}  
\label{syscalls}
\end{table}

\begin{figure*}
\centering
\includegraphics[scale=.40]{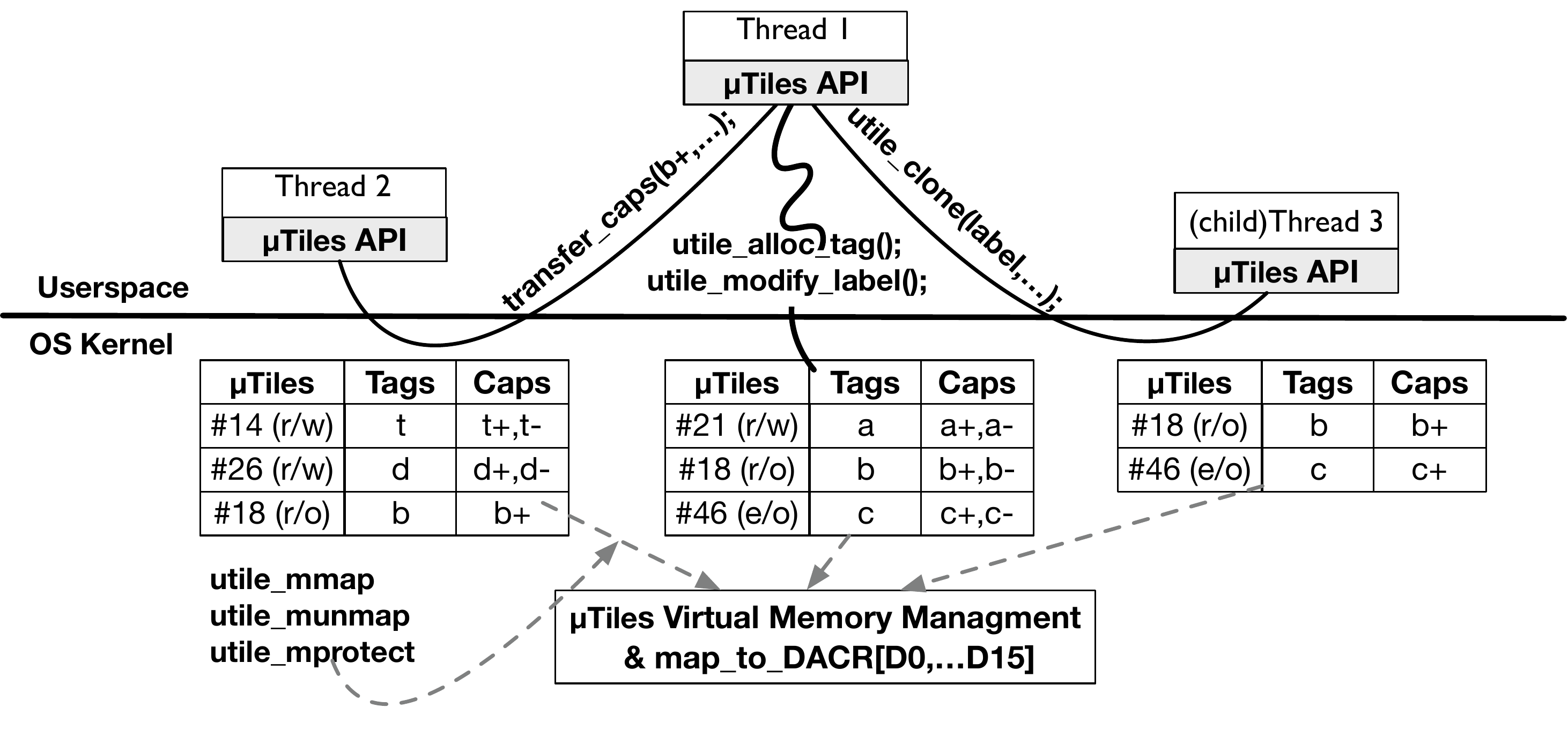}
\caption{\textmu Tiles threading abstraction: each thread is a security principal, it can define security policies for controlling its \textmu Tiles collection, and pass its capabilities to other threads. The kernel enforces the security policies and handles virtual memory management of \textmu Tiles.}
\label{threading}
\end{figure*}

\subsection{\textmu Tiles Threads}\label{labeling}
Each thread may have multiple \textmu Tiles attached to it. There is no concept of inheriting credentials and capabilities by default (e.g., in the style of \texttt{fork}) as this makes reasoning about security difficult~\cite{baumann2019fork}. For a \textmu Tile to propagate, it must be through transferring capabilities; this can be done directly by calling \texttt{utile\_transfer\_caps} for ``plus'' capabilities and \texttt{utile\_grant} for declassification or endorsement. Both these operations are also possible via specific arguments of \texttt{utile\_clone} syscall when creating a child thread. Figure~\ref{threading} shows how each thread can use the \textmu Tiles API for creating tags, changing labels, and passing capabilities to other threads. 
For instance, thread $2$ gains access to \textmu Tile $18$ by directly getting the $b^{+}$ capability from thread $1$. Since it does not have the $b^{-}$ capability, it cannot change \textmu Tile $18$ permissions or its memory mappings. 

It should be noted that \textmu Tile ID is not the same as its label.
All security-critical data structures for manging labels are stored inside the kernel, so they can not be modified by userspace attackers.
Table~\ref{syscalls} describes the userspace \textmu Tile API.
Threads can lock access or permission changes of their \textmu Tiles via \texttt{utile\_lock}, which temporarily change \textmu Tile tag to restrict any modifications of \textmu Tiles state. A locked \textmu Tile can only be accessed by calling \texttt{utile\_unlock}.\\
A tagged thread can create a child by calling \texttt{utile\_clone}; the child thread does not inherit any of its parent's capabilities. However, the parent can create a child with a list of its \textmu Tiles and selected capabilities as an argument of \texttt{utile\_clone}. For instance, in Figure~\ref{threading}, thread $1$ creates its child with only a ``plus'' capability to two of its \textmu Tiles ($18, 46$).

\subsubsection{\textmu Tiles Memory Management}

\textmu Tiles dedicated VM abstraction provides a familiar semantics for \textmu Tiles-aware memory management, VM tagging, mappings, protection, page faults handling, and least privilege enforcement.
It bypasses most of the kernel's paging abstraction. Hence, it does not require extensive modifications to the kernel memory management structures that might otherwise introduce security holes due to inevitable TLB and memory management bugs~\cite{tlbbug}. Threads' security policy enforcement is done by adding custom security hooks in the VM interfaces that check the correct flow of labels (\S\ref{labeling}).\\
To improve performance (\cref{dp}), the VM abstraction maps per thread's high-level security policies and memory management interface to the underlying hardware domains that also hide its limitations(\S\ref{challenges}). 
Example code~\ref{api} shows a basic way of using \textmu Tiles to protect sensitive content in a single thread. 

An application creates a new \textmu Tile by calling \texttt{utile\_create}; the kernel creates a unique tag with both capabilities (since it is the owner) and adds it to the thread's label and capability lists, and returns a unique ID. Then the owner thread maps pages to its \textmu Tile by calling \texttt{utile\_mmap} that updates the \textmu Tile's metadata with its address space ranges. The kernel allows mappings based on the thread's labels and free hardware domains. If there is a free hardware domain, it maps pages to that domain and places it to \textmu Tiles cache. When the \textmu Tiles already exists in the cache, further access to it is fast. When there is no free hardware domain, we have to evict one of the \textmu Tiles from the cache and map the new \textmu Tile metadata to the freed hardware domain; this requires storing all the necessary information for restoring the evicted \textmu Tile, such as its permission, address space range, and label. The caching process can be further optimized by tuning the eviction rate and suitable caching policies similar to libMPK~\cite{park2018libmpk}.

\begin{lstlisting}[caption={Basic \textmu Tiles usage},captionpos=b,label={api}]
    /* create a utile */
    int utile_id = utile_create(); 
    
    /* map a memory region to the utile */
    memblock = (char*) utile_mmap(utile_id, addr, len, prot , 0, 0); //
    
    // set permissions by utile_mprotect

    /* allocate memory from utile */
    private_blk = (char*) utile_malloc(utile_id, priv_len);

    /* make utile inaccessible */
    lock_utile(utile_id);

    //... untrusted computations ....//

    /* make utile accessible */
    unlock_utile(utile_id);

    //... trusted computations ....//

    /* cleanup utile */
    utile_free(private_blk);
    utile_munmap(utile_id, memblock,len);


\end{lstlisting}

The application uses \texttt{utile\_malloc} and the \textmu Tile ID to allocate memory within the \textmu Tile boundaries (utile\_malloc), and \texttt{utile\_free} to deallocate memory or \texttt{utile\_mprotect} to change its permissions; Table~\ref{umem} shows the familiar API for \textmu Tiles memory management. 
To mitigate attacks inside a single thread, unauthorized access to  \textmu Tile, by accident or other malicious code, are restricted once the owner calls \texttt{utile\_lock}. Then application developer can allow only her trusted functions or necessary parts of the code to gain access by calling \texttt{utile\_unlock}. For example, our single-threaded OpenSSL uses this mechanism for isolating private keys from vulnerabilities like Heartbleed bug~\cref{openssl}). \\
The VM manager has a separate fault handler for \textmu Tiles-specific cases. Illegal access to \textmu Tiles causes domain faults that our handler logs (e.g., violating thread information) and terminates it with a signal.

   \begin{table}[t]
\resizebox{\columnwidth}{!}{%

    \begin{tabular}{|l|l|}
    \hline
     Name    & Description                                 \\ \hline
    utile\_create$\to{id}$    & Create a new \textmu Tile                       \\ \hline
    utile\_kill($id$)    & Destroy a \textmu Tile                       \\ \hline
    utile\_malloc($id,size$)$\to${void*}    & Allocate memory within a \textmu Tile                       \\ \hline
    utile\_free($id,void*$)     & free memory from a  \textmu Tile                        \\ \hline
    utile\_mprotect($id,...$) & change an \textmu Tile's pages permission      \\ \hline
    utile\_mmap($id,...$)$\to${void*} & Map a page group to a \textmu Tile      \\ \hline
    utile\_munmap($id,...$) & Unmap all pages of a \textmu Tile     \\ \hline
    utile\_get($id$)$\to${$perms$}     & Get a \textmu Tile permission      \\ \hline
    \end{tabular}}
    \caption{Some of userspace \textmu Tiles memory management API. Each \textmu Tile has an $id$ and is a tagged kernel object internally. \textmu Tiles access control is checked within the kernel.  }
    \label{umem}
\end{table}

\

\section{Implementation}\label{imp}

\textbf{\textmu Tiles Kernel Modifications:}
The \textmu Tiles access control and the security model is implemented as a new Linux Security Module (LSM)~\cite{morris2002linux} with only four custom hooks. The LSM initializes the required data structures, such as the label registry. Access control system calls (Table~\ref{syscalls}) for enforcing least privilege are implemented as a part of the LSM, including locking \textmu Tiles, transferring capabilities, authority operations, and declassification based on the labeling model we described(\S\ref{labeling}). 

We modify the Linux task structure to store the metadata required to distinguish \textmu Tiles threads from regular ones.
Specifically, we add fields for storing \textmu Tiles metadata, label/ownership as an array data structure holding its tags (each tag is a 32-bit identification whose upper 2 bits stores plus and minus capabilities), a capability list; all included as a specific task's \texttt{cred->security} data structure. We implemented a hash table-based registry to make mostly used operations (e.g., store, set, get, remove) on these data structures more efficiently.

 The LSM also provides custom security hooks for parsing userspace labels to the kernel (\texttt{copy\_user\_label}), labeling a task (\texttt{set\_task\_label}), checking whether the task is labeled (\texttt{is\_task\_labeled}), and checking if the information flow between two tasks is allowed (\texttt{check\_labels\_allowed}). These security hooks are added in various places within the kernel to guard \textmu Tiles against unauthorized access or permission change by either the POSIX API (e.g., mmap, mprotect, fork) or the \textmu Tiles API. For example, forking a labeled task should not copy its labels and capability lists, and this is enforced using these security hooks. As another example, \textmu Tiles-independent applications that using traditional POSIX API can not perform any unauthorized memory allocation from a random \textmu Tile or mapping pages to it; this is restricted via the security hooks that are placed in the kernel's virtual memory management layer similar to the \textmu Tiles VM manager (Table~\ref{umem}).
 
The \textmu Tiles virtual memory abstraction is implemented as a set of kernel functions similar to their Linux equivalents (e.g., \texttt{do\_mmap}, \texttt{do\_munmap} and \texttt{do\_mprotect}) with similar high-level semantics but replaces the paging compexity with simpler hardware domain-based operations.
 When an application creates a \textmu Tile by calling \texttt{utile\_create} and maps an address range to it via \texttt{utile\_mmap}, The \textmu Tiles VM manager tags a 1MB aligned address space that covers the requested range, stores \textmu Tiles metadata, maps it to a free hardware domain and updates the \textmu Tiles cache.\\
When \textmu Tiles are mapped to hardware domains, the exact physical domain number is hidden from the userspace code to avoid possible misuse of the API. The mappings between \textmu Tiles and hardware domains are maintained through a cache-like structure similar to libmpk~\cite{park2018libmpk}.  A \textmu Tile is inside the cache if it is already associated with a hardware domain; otherwise, it evicts another \textmu Tile based on the least recently used (LRU) caching policy while saving all require metadata for restoring the \textmu Tile mapping and permission flags. 

\textmu Tiles owners (or authorities) can change their \textmu Tiles' permission via \texttt{utile\_mprotect}. This operation is faster when the requested permission matches one of the domain's supported options (Table ~\ref{domains}) or undergo the overhead of effecting TLB.
Any violation of \textmu Tiles permissions causes a \textmu Tiles fault that leads to the violating thread being terminated. 

\textbf{Userspace:}
To reduce the size of the TCB, we did not modify existing system libraries (e.g., glibc) and instead provided a small userspace library for \textmu Tiles operations that summarized in Tables~\ref{syscalls} and ~\ref{umem}.
As demonstrated, the library supports a familiar API for memory management within a \textmu Tile, including \texttt{utile\_malloc} and \texttt{utile\_free} for memory management; which is implemented as a custom memory allocator similar to HeapLayer~\cite{berger2001composing}.This allocates memory from an already mapped \textmu Tile. For each \textmu Tile, there is a memory domain in-kernel metadata structure that keeps essential information such as the \textmu Tile address space range (base and length) and the two lists of free blocks from the head and tails of the \textmu Tile region that is used when searching for free blocks of memory.

\section{Evaluation}\label{eval}

We evaluated our implementation of \textmu Tiles on a Raspberry Pi 3 Model B~\cite{rpi3} that uses a Broadcom BCM2837 SoC with a 1.2 GHz 64-bit quad-core ARM Cortex-A53 processor with $32KB$ L1 and $512KB$ L2 cache memory, running a 32-bit unmodified Linux kernel version 4.19.42 and glibc version 2.28 as the baseline.
We use microbenchmarks and compartmentalize real-world applications to evaluate \textmu Tiles in terms of performance and usability (\cref{dp} and \cref{challenges}) by answering the following questions: 

\begin{itemize}[noitemsep]
    \item What is the initialization and runtime overhead of \textmu Tiles? How does utilizing hardware domains impact performance?

    \item Are \textmu Tiles practical and adaptable for real-world applications? How much application change and programming effort is required? What is the performance impact? How does it perform for hardening a multi-threaded environment?

   \item What is the memory footprint of \textmu Tiles? Is it suitable for small IoT devices? How much memory does it add (statically and dynamically) to both the kernel and userspace?

\end{itemize}

\subsection{Microbenchmarks}

 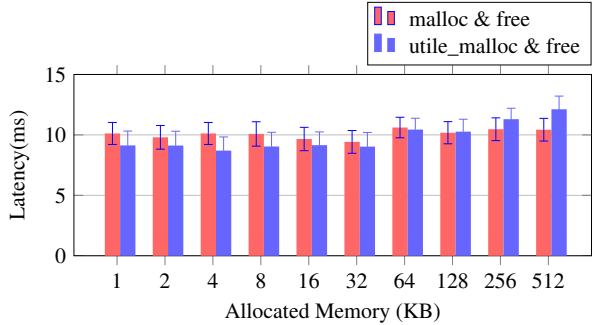
\begin{figure}[t]
\resizebox{\columnwidth}{!}{%
 \begin{tikzpicture}
    \begin{axis}[
        width=0.6\textwidth,
        height=4.5cm,
        bar width=7pt,
        % commented to see, if the result is correct
%        major x tick style=transparent,
        symbolic x coords={1,2,4,8,16,32,64,128,256,512},
        xlabel={Allocated Memory (KB)},
        xtick=data,
        %x label style={at={(axis description cs:0.5,-0.1)},anchor=north},
        % changed from `\pgflinewidth' to 1 easier see that the result is correct
        ymin=0,
         ymax=15,
        % changed to 0 to make the computation of the shift easier
        ybar=0pt,
        ymajorgrids=true,
        ylabel={Latency(ms)},
        legend cell align=left,
        legend style={
            at={(1,1.05)},
            anchor=south east,
            column sep=1ex
        },
    ]
        % add a scope around the to shift bars
        \begin{scope}[
            % shift the bars accordingly
            xshift={0.1*\pgfplotbarwidth},
            % and just "fill" the bars
            % (to avoid overlapping of the more right bars to the more left ones)
            draw=none,
        ]
            % your style could heavily be simplified to just providing the `+'
            % sign and the color as option
            %\addplot+ [bblue]  coordinates
             %   {(St.~2,10006) (St.~3,99895) (St.~4, 99867)};
            %\addplot+ [rred]   coordinates
             %   {(St.~1,1) (St.~2,10006) (St.~3,99448) (St.~4, 99487)};

\addplot+ [draw=none, fill=red!60,
            error bars/.cd,
                y dir=both,
                % (changed from `y explicit` so the error bars are (clearly) visible
                y explicit relative,
        ] coordinates {
            (1,10.12) +- (0,0.09)
            (2,9.80) +- (0,0.1)
            (4,10.12) +- (0,0.09)
            (8,10.08) +- (0,0.1)
            (16,9.66) +- (0,0.1)
            (32,9.42) +- (0,0.1)
            (64,10.61) +- (0,0.08)
            (128,10.18)  +- (0,0.09)
            (256,10.47) +- (0,0.09)
            (512,10.43) +- (0,0.09)
        };
        
        \addplot+ [draw=none, blue!60,
            error bars/.cd,
                y dir=both,
                % (changed from `y explicit` so the error bars are (clearly) visible
                y explicit relative,
        ] coordinates {          
            (1,9.13) +- (0,0.13)
            (2,9.12) +- (0,0.13)
            (4,8.7) +- (0,0.13)
            (8,9.04) +- (0,0.13)
            (16,9.15) +- (0,0.12)
            (32,9.03) +- (0,0.13)
            (64,10.44) +- (0,0.09)
            (128,10.27)  +- (0,0.1)
            (256,11.30) +- (0,0.08)
            (512,12.12) +- (0,0.09)
                        };

        \end{scope}

\legend{malloc \& free,utile\_malloc \& free}
    \end{axis}
\end{tikzpicture}}
\caption{Cost of \textmu Tiles memory allocation (malloc \& free). On average \texttt{utile\_malloc} outperforms \texttt{malloc} by a small rate ($0.03\%$).}

\label{memfig}
 \end{figure}

\textbf{Creating \textmu Tiles:}
Table~\ref{mmap} tests the cost of creating and mapping pages to \textmu Tiles using \texttt{utile\_mmap} when \textmu Tiles are directly mapped to hardware domains, as compared to virtualized \textmu Tiles when there is no free hardware domain and requires evicting \textmu Tiles from the cache. The results show that when there is a free hardware domain, the performance improves by $4.9\%$ compare to the virtualized one. Note that creating \textmu Tiles is usually a one-time operation at the initial phase of an application.

  \begin{table}[H]
 \resizebox{\columnwidth}{!}{%

    \begin{tabular}{|l|l|l|}
    \hline
    Operation                                           & Overhead & stddev    \\ \hline
    Direct utile\_mmap/munmap   & 4.8\%                        & +- 0.17\% \\ \hline
    Virtualised utile\_mmap/munmap & 10.01\%                      & +- 0.15\% \\ \hline
    \end{tabular}}
    \caption{Cost of creating \textmu Tiles when directly mapped to hardware domains vs virtualized mapping that requires \textmu Tiles caching. The results show the average of 10000 runs.}
    \label{mmap}
\end{table}

\textbf{Memory protection \& allocation:}
Changing \textmu Tile permissions and memory allocation operations inside \textmu Tiles have the most impact on runtime overhead. We evaluated the performance comparison of \texttt{utile\_mprotect} vs glibc \texttt{mprotect} based on permission flags. Since hardware memory domains do not have flexible access control options (\cref{challenges}), we cannot benefit from a control switch of domains using the DACR register for all possible permission flags such as the RO, WO, and EO variants. Our results show that on average \texttt{utile\_mprotect} is $1.17$x faster than \texttt{mprotect} for no access (PROT\_NONE) or RW permissions (PROT\_READ | PROT\_WRITE), but $1.3$x slower for read/write/execute-only options that are emulated.

Allocating memory using \texttt{utile\_malloc} is on average 1.08x faster than glibc \texttt{malloc} for blocks $ \leq64KB$ and introduces a small overhead ($8.3$\%) for larger blocks ($>64KB$) as demonstrated in Figure~\ref{memfig}. This cost can be optimized by using high-performance memory allocators~\cite{lietar2019snmalloc}. We report the average of running microbenchmarks 20000 times and show
how utilizing \textmu Tiles provides small overhead for memory allocation and permission changes.

\begin{figure}[t]
\resizebox{\columnwidth}{!}{%

 \begin{tikzpicture}
    \begin{axis}[
        width=0.6\textwidth,
        height=4.5cm,
        bar width=7pt,
        % commented to see, if the result is correct
%        major x tick style=transparent,
        symbolic x coords={utile-clone,pthread,fork},
        xlabel={},
        xtick=data,
        %x label style={at={(axis description cs:0.5,-0.1)},anchor=north},
        % changed from `\pgflinewidth' to 1 easier see that the result is correct
        ymin=1,
        % changed to 0 to make the computation of the shift easier
        ybar=0pt,
        ymajorgrids=true,
        ylabel={Latency(ms)},
        ymode=log,
         legend cell align=left,
        legend style={
            at={(1,1.05)},
            anchor=south east,
            legend columns=2,
            column sep=1ex
        },
    ]
        % add a scope around the to shift bars
        \begin{scope}[
            % shift the bars accordingly
            % and just "fill" the bars
            % (to avoid overlapping of the more right bars to the more left ones)
            draw=none,
        ]

        \addplot [draw=none, fill=red!60] coordinates {
        (utile-clone,56.099) 
        (pthread,60.439)
        (fork,479.231) };
         \addplot [draw=none, fill=green!50] coordinates {
          (utile-clone,60.167) 
          (pthread,62.460)
          (fork,477.621) };
          
         \addplot [draw=none, fill=blue!50] coordinates {
         (utile-clone,3.688) 
         (pthread,3.914)
         (fork,2143) };

           \addplot [draw=none, fill=blue!90] coordinates {
           (utile-clone,3.941) 
           (pthread,3.731)
           (fork,1884) };

        \end{scope}

\legend{Lunch(1MB),Lunch(2MB),Join(1MB), Join(2MB) }
    \end{axis}
\end{tikzpicture}}
\caption{Overhead of creating \textmu Tiles-enabled threads: the results are the average of 100000 runs with 1MB and 2MB heap sizes. On average, \texttt{utile\_clone} latency is $5.39\%$ lower than of \texttt{pthread\_create}. }
\label{fork}
 \end{figure}
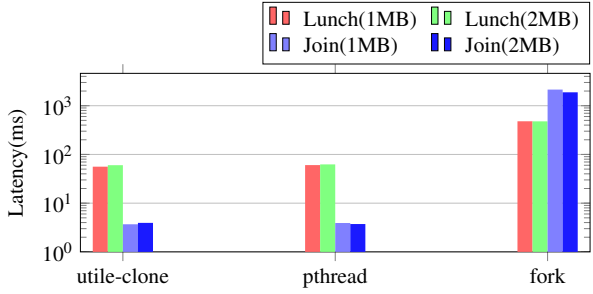

\textbf{Threading:}
We tested the cost of \textmu Tile threading operations (creating and joining) through \texttt{utile\_clone} that creates \textmu Tile-aware threads; \texttt{utile\_clone} internally uses the \texttt{clone} syscall with minor modifications to restrict any credential sharing with the child by default (instead it provides additional clone options for passing parent's capabilities to its child). We implemented \texttt{utile\_join} similar to \texttt{waitpid}. Table~\ref{fork} shows \texttt{utile\_clone} outperforms \texttt{pthread\_create} by $0.56\%$ and \texttt{fork} by $83.01\% $.  This gain is attributed to the \texttt{utile\_clone} simply doing less sharing for initializing new threads.

\textbf{Codebase overhead:}
Another factor towards the usability of \textmu Tiles is the codebase size, which is important both from a security perspective and the resource limitations of small IoT devices. We implemented \textmu Tiles as a Linux kernel patch with no dependency on any third-party library. As Table~\ref{totalloc} shows, it adds less than $5.5K$ LoC in total to both the kernel ($\approx 3K$) and userspace ($< 2.5K$).  It adds $7KB$ to the kernel image size and adds $204KB$ for kernel slabs at runtime. The userspace library only needs $\approx 10KB$ of memory. These results show that the \textmu Tiles memory footprint is extremely low and suitable for many resource-constrained uses.

   \begin{table}[htb]
\resizebox{\columnwidth}{!}{%

    \begin{tabular}{|l|l|l|}
    \hline
    Overhead                & Linux Kernel               & Userspace     \\ \hline
    Added LoC             & 3023                       & 2405          \\ \hline
    Static Memory footprint & static(7KB) slab(204KB) & Static(10KB)  \\ \hline
    \end{tabular}}
    \caption{\textmu Tiles codebase size and Memory footprint in the Linux Kernel and userspace}
    \label{totalloc}
\end{table}
 
 \subsection{OpenSSL}\label{openssl}
 
    \begin{figure}[t]
\resizebox{\columnwidth}{!}{%

 \begin{tikzpicture}
    \begin{axis}[
        width=0.7\textwidth,
        height=4.5cm,
        bar width=7pt,
        % commented to see, if the result is correct
%        major x tick style=transparent,
        symbolic x coords={1,2,4,8,16,32,64,128,256, 512},
        xlabel={number of requests},
        xtick=data,
        %x label style={at={(axis description cs:0.5,-0.1)},anchor=north},
        % changed from `\pgflinewidth' to 1 easier see that the result is correct
        ymin=1,
        % changed to 0 to make the computation of the shift easier
        ybar=0pt,
        ymajorgrids=true,
        ylabel={Latency(s)},
        ymode=log,
        legend cell align=left,
        legend style={
            at={(1,1.05)},
            anchor=south east,
            column sep=1ex
        },
    ]
        % add a scope around the to shift bars
        \begin{scope}[
            % shift the bars accordingly
            % and just "fill" the bars
            % (to avoid overlapping of the more right bars to the more left ones)
            draw=none,
        ]

\addplot  [draw=none, fill=blue!40] coordinates { (1,0.618) (2,1.396) (4,2.932) (8,5.260) (16,10.794) (32,22.569) (64,44.525) (128,95.875) (256,180.768) (512,365.445)  };
\addplot  [draw=none, fill=green!60] coordinates { (1,0.750) (2,1.442) (4,2.694) (8,5.371) (16,10.627) (32,21.275) (64,45.495) (128,94.714)(256, 188.252) (512,370.352)  };
\addplot  [draw=none, fill=red!60] coordinates { (1,0.782) (2,1.278) (4,2.776) (8,5.246) (16,10.451) (32,22.861) (64,49.201 ) (128,98.023) (256, 209.613)(512,386.236)  };

        \end{scope}

\legend{original,\textmu Tiles (single \textmu Tile), \textmu Tiles (per session \textmu Tiles)}
    \end{axis}
\end{tikzpicture}}
\caption{Overhead of httpd on unmodified OpenSSL vs \textmu Tiles-enabled one.}
%overhead of using single \textmu Tile for protecting private keys is $0.47\%$ while using multiple \textmu Tiles (one per session) causes a larger overhead of $3.67\%$ compare to using unmodified OpenSSL baseline.}
\label{httpd}
 \end{figure}
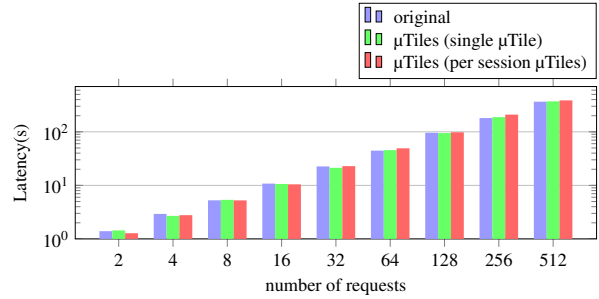
 
OpenSSL is a widely used open-source library implementing cryptography operations and the transport layer security (TLS) protocol. It handles sensitive content such as private keys and encrypted data. Hence it significantly benefits from isolating its sensitive content in separate compartments to mitigate information leakage attacks (e.g., Heartbleed). We modified OpenSSL to utilize \textmu Tiles for protecting private keys from potential information leakage by storing the keys in protected memory pages inside a single \textmu Tile or multiple \textmu Tiles assigned per private key. Using multiple \textmu Tiles provides stronger security while adding more overhead due to the cost of caching \textmu Tiles.

To enable \textmu Tiles inside OpenSSL, all the data structures that store private keys such as \texttt{EVP\_PKEY} needed protected heap memory allocation. This meant replacing \texttt{OpenSSL\_malloc} wit \texttt{utile\_malloc} and using \texttt{utile\_mmap} at the initialization phase for creating one or multiple (per session) \textmu Tiles to store private keys. After storing the keys, access to \textmu Tiles is disabled by calling \texttt{utile\_lock}. Only trusted functions that require access to private keys (e.g., \texttt{EVP\_EncryptUpdate} or \texttt{pkey\_rsa\_encrypt/decrypt}) can access \textmu Tiles by calling \texttt{utile\_unlock}. Modifying OpenSSL required fairly small code changes, and added only $281$ LoC.

%    \begin{tabular}{|l|l|l|}
%    \hline
%    Overhead                & OpenSSL               & LevelDB     \\ %\hline
%    Added LoC               & 281                       & 157          \\ \hline
%    \end{tabular}
%    \caption{Programming effort of enabling \textmu Tiles in OpenSSL and %LevelDB}

We measured the performance overhead of \textmu Tiles-enabled OpenSSL by evaluating it on the Apache HTTP server (httpd) that uses OpenSSL to implement HTTPS. Table~\ref{httpd} shows the overhead of ApacheBench httpd with both the original OpenSSL library and the secured one with \textmu Tiles. ApacheBench is launched 100 times with various request parameters. We choose the TLS1.2 DHE-RSA-AES256-GCM-SHA384 algorithm with 2048-bit keys as a cipher suite in the evaluation.

The results show that on average \textmu Tiles introduces $0.47\%$  performance overhead in terms of latency when using a single \textmu Tile for protecting all keys, and $3.67\%$ overhead when using a separate \textmu Tile per session key. In the single \textmu Tile case, the negligible overhead is mainly caused by in-kernel data structure maintenance for enforcing privilage separation and handling \textmu Tiles metadata. In the multiple-\textmu Tiles case, since httpd utilizes more than 16 \textmu Tiles (allocates a new \textmu Tile per session), it causes higher overhead due to the caching costs within the kernel.

\subsection{LevelDB}\label{leveldb} 

To show how \textmu Tiles can be used for hardening multi-threaded applications, we modified Google's LevelDB that is a fast key-value store and storage engine used by many applications as a backend database. It supports multithreading for both concurrent writers to insert data into the database as well as concurrent read to improve its performance. However, there is no privilege separation between threads, so threads can not communicate securely with the database and protect their private content from other threads. We modified LevelDB to evaluate performance overhead of using the \textmu Tiles secure threading model when each thread has its own private storage that cannot be accessed by other threads.

We replaced the LevelDB threading backend (\texttt{env\_posix}) that uses \texttt{pthreads} with \textmu Tiles-aware threading, where each thread creates an isolated \textmu Tile to protect its private storage and sensitive computations. We used the LevelDB \texttt{db\_bench} tool (without modification) for measuring the performance overhead of \textmu Tiles.

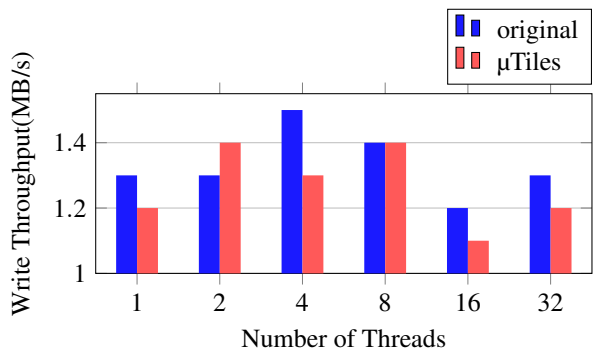
\begin{figure}[htb]
\resizebox{\columnwidth}{!}{%

 \begin{tikzpicture}
    \begin{axis}[
        width=0.5\textwidth,
        height=4cm,
        bar width=8pt,
        % commented to see, if the result is correct
%        major x tick style=transparent,
        symbolic x coords={1,2,4,8,16,32},
        xlabel={Number of Threads},
        xtick=data,
        %x label style={at={(axis description cs:0.5,-0.1)},anchor=north},
        % changed from `\pgflinewidth' to 1 easier see that the result is correct
        ymin=1,
        % changed to 0 to make the computation of the shift easier
        ybar=0pt,
        ymajorgrids=true,
        ylabel={Write Throughput(MB/s)},
        legend cell align=left,
        legend style={
            at={(1,1.05)},
            anchor=south east,
            column sep=1ex
        },
    ]
        % add a scope around the to shift bars
        \begin{scope}[
            % shift the bars accordingly
            % and just "fill" the bars
            % (to avoid overlapping of the more right bars to the more left ones)
            draw=none,
        ]

\addplot  [draw=none, fill=blue!90] coordinates { (1,1.3) (2,1.3) (4,1.5) (8,1.4) (16,1.2) (32,1.3 )   };
\addplot  [draw=none, fill=red!65 ] coordinates { (1,1.2) (2,1.4) (4,1.3) (8,1.4) (16,1.1) (32, 1.2) };

        \end{scope}

\legend{original, \textmu Tiles}
    \end{axis}
\end{tikzpicture}}
\caption{LevelDB: performance overhead of \textmu Tiles-based multithreading compare to pthread-based in terms of write throughput ($5\%$).}
\label{ldb1}
 \end{figure}

\begin{figure}[htb]
\resizebox{\columnwidth}{!}{%

 \begin{tikzpicture}
    \begin{axis}[
        width=0.5\textwidth,
        height=4cm,
        bar width=8pt,
        % commented to see, if the result is correct
%        major x tick style=transparent,
        symbolic x coords={1,2,4,8,16,32},
        xlabel={Number of Threads},
        xtick=data,
        %x label style={at={(axis description cs:0.5,-0.1)},anchor=north},
        % changed from `\pgflinewidth' to 1 easier see that the result is correct
        ymin=1,
        % changed to 0 to make the computation of the shift easier
        ybar=0pt,
        ymajorgrids=true,
        ylabel={Read Throughput(MB/s)},
        legend cell align=left,
        legend style={
            at={(1,1.05)},
            anchor=south east,
            column sep=1ex
        },
    ]
        % add a scope around the to shift bars
        \begin{scope}[
            % shift the bars accordingly
            % and just "fill" the bars
            % (to avoid overlapping of the more right bars to the more left ones)
            draw=none,
        ]

\addplot  [draw=none, fill=blue!45] coordinates { (1,78.4) (2,80.2) (4,150.3) (8,268.5) (16,300.5) (32,280.6)  };

\addplot  [draw=none, fill=green!90] coordinates { (1,80.4) (2,79.0) (4,153.3) (8,299.3) (16,301.6) (32, 267.9)};

        \end{scope}

\legend{original, \textmu Tiles}
    \end{axis}
\end{tikzpicture}}
\caption{LevelDB: performance overhead of \textmu Tiles-based multithreading compare to pthread-based in terms of read throughput ($1.98\%$).}
\label{ldb2}
 \end{figure}
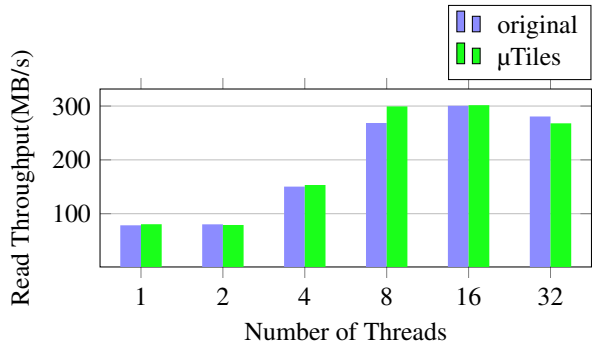

We generate a database with 400K records with 16-byte keys and 100-byte values (a raw size of 44.3MB). The number of reader threads is set to 1, 2, 4, 8, 16,  and 32 threads for each successive run. The threads operate on randomly selected records in the database. The results in Figures~\ref{ldb1} and ~\ref{ldb2} show how multithreading can improve the performance of LevelDB, and utilizing \textmu Tiles adds a small overhead on write ($5\%$) and read ($1.98\%$) throughput.  As with OpenSSL previously, modifying LevelDB required only adding $157$ lines-of-code around the codebase.

  
\section{Discussion}\label{diss}
We have shown that \textmu Tiles provides a practical and efficient mechanism for intra-process isolation and inter-thread privilege separation on data objects. However, the mechanism can still be taken further.

\subsection{Address Space Protection Limitations}

For single-threaded scenarios (e.g., event-driven servers), although \textmu Tiles can protect sensitive content from unsafe libraries or untrusted parts of the applications, it can be vulnerable if the untrusted modules are also \textmu Tiles-aware and already use the \textmu Tiles APIs.  The untrusted library can use \texttt{utile\_get} to query \textmu Tile IDs and use the API to reach them. It should be noted that this is not an issue for untrusted legacy libraries. Additionally, to remove the possibility of such attacks, it is better to run these unsafe libraries in a separate thread, which is isolated through \textmu Tiles abstraction. 

Various covert attacks~\cite{sigurbjarnarson2018nickel} and side-channel attacks such as Meltdown~\cite{lipp2018meltdown} and Spectre~\cite{kocher2018spectre} demonstrate how hardware and kernel isolation can be bypassed~\cite{hunt2019isolation}. \textmu Tiles are currently vulnerable to these class of attacks, although the existing countermeasures within the Linux kernel are sufficient protection. We believe these types of attacks are important security threats, and hardening \textmu Tiles against them could be significant future work.

\subsection{Compatibility Limitations}

Providing a solution that is compatible with various operating systems and heterogeneous hardware is challenging. Though we picked our base kernel on Linux and built the abstraction with minimal dependencies, some application modification is still required.  We believe that building more compatibility layers into our existing userspace implementation is possible. We are open-sourcing our code with further feedback and patches from the relevant upstream projects we have modified.

Although Linux is the most widespread general-purpose kernel for embedded devices, many even smaller devices depend on operating systems such as FreeRTOS.  These often use ARM Cortex-M based hardware features for isolation (such as memory protection units (MPUs)~\cite{mpu,cmsis}), or more recent CPUs with memory tagging extension~\cite{mte}. We plan to explore the implementation of the \textmu Tiles kernel memory management on these single-address space operating systems, as well as broadening the port to Intel architectures on Linux (where the memory domains support is generally simpler to use than on ARM).

\section{Related work}\label{related}
There are many software or hardware-based techniques for providing process and intra-process memory protection.  

\paragraph{OS/hypervisor-based solutions:}
Hardware virtualization features are used for in-process data encapsulation by Dune~\cite{belay2012dune} by using the Intel VT-x virtualization extensions to isolate compartments within user processes. However, overall, the overheads of such hardware virtualization-based encapsulation are much more heavy-weight than \textmu Tiles, and not practical for IoT applications. 

ERIM~\cite{vahldiek2019erim}, light-weight contexts (lwCs)~\cite{litton2016light} and secure memory views (SMVs)~\cite{hsu2016enforcing} all provide in-process memory isolation and have reduced the overhead of sensitive data encapsulation on x86 platforms.
The \textmu Tiles abstraction provides stronger security guarantees and privilege separation. It allows more flexible ways of defining security policies for legacy code -- e.g., within a single thread as in our OpenSSL example. Its small memory footprint makes it suitable for smaller devices, and it takes advantage of efficient virtual memory tagging by using hardware domains to reduce overhead.

Burow et al.~\cite{burow2019sok} also leverage the Intel MPK and memory protection extensions (MPX) to isolate the shadow stack. Our efforts to provide an OS abstraction for in-process memory protection is orthogonal but more general than these studies, which all have potential use cases for \textmu Tiles.  Our focus has also been on lowering the resource cost to work well on embedded and IoT devices, while these projects are also currently x86-only.

HiStar~\cite{zeldovich2006making} is a DIFC-based OS that supports fine-grained in-process address space isolation. It influenced our work, but we focused on providing a more general-purpose solution for small devices by basing our work on the Linux kernel instead of a custom operating system. Other DIFC-based systems only support per-process protection with very large overhead~\cite{wang2015between, krohn2007information} or need specific programming language support~\cite{roy2009laminar}.

\paragraph{Compiler \& Language Runtime:}
Various compiler techniques introduce memory isolation as part of a memory-safe programming language. These approaches are fine-grained and efficient if the checks can be done statically~\cite{elliott2018checked}. However, such isolation is language-specific, relies on the compiler and runtime, and not effective when applications are co-linked with libraries written in unsafe languages and libraries. \textmu Tiles abstractions are fine-grained enough to be useful to these tools, for example, to isolate unsafe bindings.

Software fault isolation (SFI)~\cite{wahbe1994efficient,sehr2010adapting} uses runtime memory access checks inserted by the compiler or by rewriting binaries to provide memory isolation in unsafe languages with substantial overhead. Bounds checks impose overhead on the execution of all components (even untrusted ones), and additional overhead is required to prevent control-flow hijacks, which could bypass the bounds checks~\cite{koning2017no}.
ARMLock~\cite{zhou2014armlock} is an SFI-based solution that offers lower overhead utilizing ARM memory domains.Similarly, Shreds~\cite{chen2016shreds} provides new programming primitives for in-process private memory support.\
\textmu Tiles also uses ARM memory domains for improving the performance of intra-process memory protection, but is a more flexible solution for intra-process privilege separation; it provides a new threading model for dynamic fine-grained access control over the address space with no dependency on a binary rewriter, specific compiler or programming language (See Table ~\ref{compare}).

\paragraph{Hardware-enforced techniques:}
A wide range of systems use hardware enclaves such as Intel's SGX~\cite{anati2013sgx} or ARM's TrustZone~\cite{tzcm} to provide a trusted execution environment for applications that against malicious kernel or hypervisor~\cite{guan2017trustshadow,frassetto2017jitguard,arnautov2016scone}.

The trust model exposed by these hardware features is very fixed, and usually results in porting monolithic codebases to execute within the enclaves.  EnclaveDom~\cite{melara2019enclavedom} utilizes Intel MPK to provide in-enclave privilege separation. \textmu Tiles provide better performance and more general solutions with no dependency on these hardware features; hence it can be used for in-enclave isolation and secure multi-threading to improves both security and performance of enclave-assisted applications.

Ultimately, dedicated hardware support for tagged memory and capabilities (e.g., ARM MTE~\cite{mte}) would be the ideal platform to run \textmu Tiles on~\cite{zeldovich2008hardware}. We are planning on building this support as future work, with a view to analyzing if the overall increase in hardware complexity offsets the resource usage in software for embedded systems.
\section{Conclusions}\label{conc}

We have presented \textmu Tiles -- an OS abstraction, a set of security primitives and APIs for protecting data objects inside a shared address space, and providing flexible privileged separation for multithreaded applications. We designed \textmu Tiles to be extremely lightweight for IoT applications, with no programming language requirements, and with a small performance overhead by utilizing efficient hardware-based memory protection that makes it practical for a variety of uses cases and security-sensitive applications.
\section{Acknowledgment}

We thank Ed Nightingale, Reuben Olinsky, and Jewell Seay for helpful discussions, and David Chisnall, Jon Crowcroft, Marno van der Maas, and Ali Varamesh for feedback on earlier drafts of this paper.

%\bibliographystyle{ACM-Reference-Format}

%-------------------------------------------------------------------------------
\bibliographystyle{plain}
\bibliography{references}

%%%%%%%%%%%%%%%%%%%%%%%%%%%%%%%%%%%%%%%%%%%%%%%%%%%%%%%%%%%%%%%%%%%%%%%%%%%%%%%%
\end{document}